\documentclass[twocolumn]{aastex631-2}

\usepackage{amsmath}

\shorttitle{Optimizing MAS CCD Readout}
\shortauthors{Lin et al.}

\begin{document}

\title{Optimizing Charge-coupled Device Readout Enabled by the Floating-Gate Amplifier}

\author[0000-0001-8967-2281]{Kenneth W. Lin}
\affiliation{Department of Astronomy, University of California, Berkeley, CA 94720, USA}
\affiliation{Lawrence Berkeley National Laboratory, One Cyclotron Rd, Berkeley, CA 94720, USA}

\author[0000-0002-9964-1005]{Abby Bault}
\affiliation{Lawrence Berkeley National Laboratory, One Cyclotron Rd, Berkeley, CA 94720, USA}

\author[0000-0003-2285-1765]{Armin Karcher}
\affiliation{Lawrence Berkeley National Laboratory, One Cyclotron Rd, Berkeley, CA 94720, USA}

\author[0000-0001-9822-6793]{Julien Guy}
\affiliation{Lawrence Berkeley National Laboratory, One Cyclotron Rd, Berkeley, CA 94720, USA}

\author[0000-0001-7218-3457]{Stephen E. Holland}
\affiliation{Lawrence Berkeley National Laboratory, One Cyclotron Rd, Berkeley, CA 94720, USA}

\author[0009-0001-4005-8771]{William F. Kolbe}
\affiliation{Lawrence Berkeley National Laboratory, One Cyclotron Rd, Berkeley, CA 94720, USA}

\author[0000-0002-3389-0586]{Peter E. Nugent}
\affiliation{Department of Astronomy, University of California, Berkeley, CA 94720, USA}
\affiliation{Lawrence Berkeley National Laboratory, One Cyclotron Rd, Berkeley, CA 94720, USA}

\begin{abstract}
Multiple-Amplifier Sensing (MAS) charge-coupled devices (CCDs) have recently been shown to be promising silicon detectors that meet noise sensitivity requirements for next generation Stage-5 spectroscopic surveys and potentially, future space-based imaging of extremely faint objects on missions such as the Habitable Worlds Observatory. Building upon the capability of the Skipper CCD to achieve deeply sub-electron noise floors, MAS CCDs utilize multiple floating-gate amplifiers along the serial register to increase the readout speed by a factor of the number of output nodes compared to a Skipper CCD. We introduce and experimentally demonstrate on a 16-channel prototype device new readout techniques that exploit the MAS CCD's floating-gate amplifiers to optimize the correlated double sampling (CDS) by resetting once per line instead of once per pixel. With this new mode, we find an optimal filter to subtract the noise from the signal during read out. We also take advantage of the MAS CCD's structure to tune the read time by independently changing integration times for the signal and reference level. Together with optimal weighted averaging of the 16 outputs, these approaches enable us to reach a sub-electron noise of 0.9 e$^-$ rms pix$^{-1}$ at 19~$\mu$s pix$^{-1}$ for a single charge measurement per pixel --- simultaneously giving a 30\% faster readout time and 10\% lower read noise compared to performance previously evaluated without these techniques.
\end{abstract}


\section{Introduction} \label{sec:intro}
Silicon charge-coupled devices (CCDs) have been prized in ground and space-based optical astronomical imaging and spectroscopy for their linearity, low-noise, quantum efficiency, and flight heritage. As the ambitious scientific drivers for planned missions such as a Stage-5 (Spec-S5) spectroscopic survey for precision cosmology or the Habitable Worlds Observatory (HWO) for direct exo-Earth detection demand ever lower noise floors down to the quantum limit, new detector architectures must be developed to meet aggressive sensitivity (e.g., $< 1$ e$^{-}$ rms pix$^{-1}$) and fast readout (e.g., $\lesssim14$ $\mu$s pix$^{-1}$) requirements \citep{2025arXiv250307923B, 2016JATIS...2d1212R,2024arXiv240719176A}. Many superconducting quantum detectors have been proposed and tested but still have technical challenges that hinder wider adoption in astronomy, including multiplexing into large pixel array formats, specialized fabrication processes, and demanding cryogenic cooling and infrared radiation shielding requirements that significantly increase engineering complexity \citep[e.g.,][]{BERGGREN2013185, 10.1117/12.3020513}. Thus, the Skipper CCD has garnered interest as a potential solution because of its photon-counting capability that also retains the favorable properties of modern scientific CCD sensors and only requires passive cooling. Skipper CCDs have already proven to be scientifically valuable in several ongoing sub-GeV direct dark matter search experiments with world-leading constraints \citep[e.g.,][]{2019PhRvL.122p1801A,2023PhRvL.130q1003A,PhysRevLett.134.011804}. 

A Skipper CCD is distinguished from a conventional CCD by its use of a floating-gate amplifier, which allows charge in the sense node to be non-destructively sampled repeatedly by shuffling charge in each pixel multiple times between the summing well and sense node, thereby overcoming the $1/f$ read noise limit \citep{Janesick_1993}. The read noise thus scales as the inverse square root of the number of measurements, or ``skips," performed for charge in a pixel. The signal-to-noise ratio improvements in resolving spectral lines enabled by sub-electron noise performance in Skipper CCDs has also recently been demonstrated on sky as part of an integral field spectrograph \citep{2024SPIE13103E..0FM}. However, because its noise performance scales inversely with the number of charge measurements per pixel, and by extension, the readout time, the Skipper CCD is not ideal for most astronomical observations where factors such as the cadence, cosmic rays, and transient targets impose restrictions on the detector dead time. 

One promising detector proposed to address this limitation is the Multiple-Amplifier Sensing (MAS) CCD, which implements multiple floating-gate amplifiers interspersed along the serial register to read out pixel charge repeatedly at a faster overall time than a Skipper CCD \citep{2023AN....34430072H,10521851}. The MAS CCD has been recently demonstrated to achieve nearly single-electron noise at a 40 kHz pixel rate and has shown encouraging in-lab characterization results for 8- and 16-channel prototype devices including $>0.9995$ inter-amplifier charge transfer efficiency and low nonlinearity of 2.5\% across a dynamic range from a few hundred to 35,000 e$^-$ \citep[e.g.,][]{2024PASP..136i5002L, 10.1117/1.JATIS.11.1.011203, 2024SPIE13103E..11B, 2024SPIE13103E..10L}. While single-electron read noise meets the projected noise requirements for a next generation spectroscopic survey and is already more than a factor of eight faster than a Skipper CCD at the same noise level, there is still a trade-off in the readout speed compared to the CCDs currently used in the Dark Energy Spectroscopic Instrument (DESI) with a 66 kHz pixel rate. 

In this paper, we introduce readout techniques enabled by the floating-gate amplifiers of the MAS CCD aimed at further reducing the noise floor with a single read per amplifier while also decreasing the required pixel readout time. A modified sense node reset scheme is implemented, additional optimization in the pixel sampling process downstream from the controller is explored, and a method for optimally averaging amplifier images on the overall noise performance is studied. We experimentally demonstrate these techniques on a prototype 16-channel MAS CCD using Hydra readout electronics, a 16-channel adaptation of the DESI front-end electronics (FEE), both of which were previously described in \citet{2024PASP..136i5002L}. These improvements show the potential of an extended iteration of the MAS CCD with double the number of channels and differential readout to satisfy or even exceed the requirements of future massively multiplexed spectroscopic surveys. 

This paper is structured as follows: Section \ref{sec:opportunities} describes the readout approaches and optimal filtering approach we explore in the context of the floating-gate amplifier, Section \ref{sec:readtime} summarizes our results using the optimal filter to improve the read noise and read time, Section \ref{optimalaveraging} describes additional efforts to improve the read noise by optimally combining the amplifiers, and Section \ref{sec:summary} summarizes our main findings and comments on implications for future work. We deploy the techniques discussed in this paper for the readout of a thick, 16-channel MAS CCD with DESI electronics. Details regarding the detector, package, electronics, and baseline performance from experimental tests are found in \citet{2024PASP..136i5002L}, where a read noise of 1.03 e$^-$ pix$^{-1}$ at a read time of 26 $\mu$s pix$^{-1}$ was reported. The experimental setup used in this paper is similar to that in \citet{2024PASP..136i5002L}, but instead of liquid nitrogen cooling, an integral Stirling cryocooler (Sunpower CryoTel\textsuperscript{\textregistered} GT) now provides the cooling power for the 140 K operating temperature of the CCD.

\section{Readout Techniques and \\Optimal Filtering} 
\label{sec:opportunities}

In typical CCD charge-to-voltage conversion, prior to transferring the pixel charge into the output stage of a floating-diffusion amplifier, the previous charge must be cleared by resetting the sense capacitor and integrating this uncertain reference level, inevitably introducing a thermodynamic kTC noise from the reset \citep{Carnes1972}. By subtracting this reference level from the signal level, the true value of the charge in each pixel can be determined while minimizing the kTC noise and eliminating the DC offset of the reference voltage. This process is known as correlated double sampling (CDS) and is performed after each pixel reset \citep{1050448,47768}.

With a floating-diffusion amplifier, charge cannot be recovered after measurement because the charge is transferred from the serial register onto an output node, a reverse biased diode which is directly connected to a MOSFET gate operated as a source-follower \citep{Theuwissen1995,KRAFT1995372}. This effectively adds the signal charge to the charge of the output node. The signal charge information is lost since if the charge were to exit, the potential under the output gate should be set equal or below the output diode, injecting charge mixed from both the diode and signal \citep{2001sccd.book.....J}. After sampling, the charge must be cleared with a reset to avoid charge being added to each successive pixel. With a floating-gate amplifier however, the signal charge is only capacitively coupled to the output node, allowing charge to be clocked away and leaving the gate exactly in the same state as it was prior to the presence of signal charge \citep{1155181,KRAFT1995372,Theuwissen1995,628823}. Therefore, the requirement for a reset per pixel is not apparent and we explore reducing the resets we perform to only once per line.

Enabled by a reset per line scheme, we can optimize the CDS process by extracting and digitizing the reference level and signal level separately as part of pre-processing performed outside of the readout electronics FPGA. Since the reset transistor is only active once per line, the same reset level is maintained for all the pixels in a line. If the reset-level values for each pixel in a line are directly extracted, this level can be filtered before the usual subtraction from the signal level is performed. This low-pass filtering serves to suppress the high-frequency noise components present in the reference level in a given line. Removing this noise thus reduces the overall noise when the CDS is performed, and extracting the reset level for signal processing prior to the subtraction from the signal level becomes potentially advantageous. A consequence of handling the reset and signal levels downstream is that the integration time set for either level can be different, offering the possibility of optimizing the integration times of each level independently and reducing the overall read time.

We separately extract the signal level $S$ and reset level $R$, then perform an optimized weighted subtraction of the reference level from the signal by summing over different pixel shifts of the reference level columns:
\begin{equation}
    \label{eq:optimal_sub}
    S^\prime_t = S_t - \sum_{i} w_i R_{t+i}.
\end{equation}
Here, $t$ is an index that denotes the relevant signal pixels, $i$ is the number of pixels the reset level is being shifted by, $w_i$ is the optimal weight for each value of $i$. The array of weights ($w_i$) form a filter of  width $W$ (the size of the array). This optimal filter is essentially a Wiener filter which is used to obtain the best estimate of a signal given the noise \citep{wiener}. A Wiener filter assumes that the signal of interest (here the reference level $R$) and its noise each have their own temporal auto-correlation function but that they are uncorrelated with each other, which is true in this setup as the noise in the reference level is the electronic noise of the measurement. However, instead of directly applying the Wiener filter, we performed a least squares minimization to determine the optimal weights $w$ that minimizes the noise in the reference level. 

The weights, $w$, are calculated using only the overscan columns of the image. We calculate the weights such that the noise in the reference level is minimized by performing a $\chi^2$ minimization:
\begin{equation}
    \label{eq:chi2}
    \chi^2 = \sum_t \left[S_t - \sum_{i} w_i R_{t+i} \right]^2.
\end{equation} 

The weights could also be calculated using the prescan columns of the image, however, we chose to use the overscan region for this calculation as the size of the prescan region limits the maximum filter width we can use. Once the weights are determined using the overscan data, the filtered reference level is subtracted from the signal for each pixel measured by an amplifier, and for each of the 16 amplifiers independently (we determine a different set of weights for each amplifier). The images from the 16 amplifiers are then combined by stacking and taking a weighted average over each of the images. The overscan columns of the combined image are then used to calculate the read noise. 

\begin{figure}
    \centering
    \includegraphics[width=\linewidth]{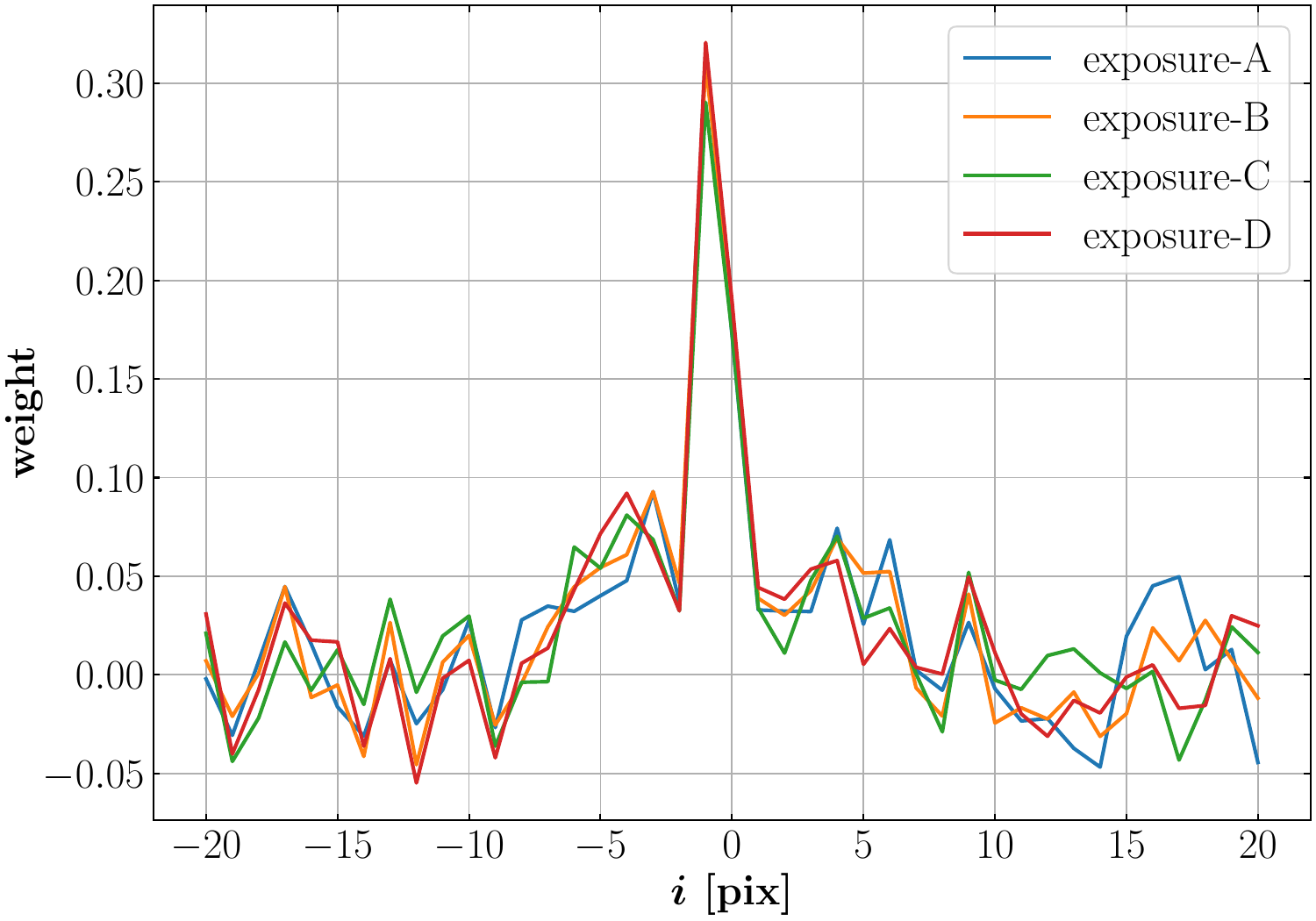}
    \caption{Weights calculated for $W = 41$. There is a peak around $i = -1$ and some oscillatory behavior in the tails, following the pattern from a Wiener filter. This oscillatory behavior is not noise, as it is consistent across multiple exposures (A,B,C,D) with the same read time of 25.4 $\mu$s pix$^{-1}$. The shape of the filter is consistent across the four exposures as well.}
    \label{fig:weights}
\end{figure}

While the filter width, $W$, can be set to any value (we discuss the optimal value later in this section), we show the calculated weights for $W = 41$ in Figure \ref{fig:weights}. The weights peak at $i=\{-1,0\}$, with some oscillations in the tails (as expected for the Wiener filter of a signal with a finite correlation length and white noise). The maximum weights at $i=\{-1,0\}$ correspond to values of the reference level taken just before and after measuring the signal. 
We attribute the difference in magnitude to the timing of each measurement not being exactly the same. 

\begin{figure}
    \centering
    \includegraphics[width=\linewidth]{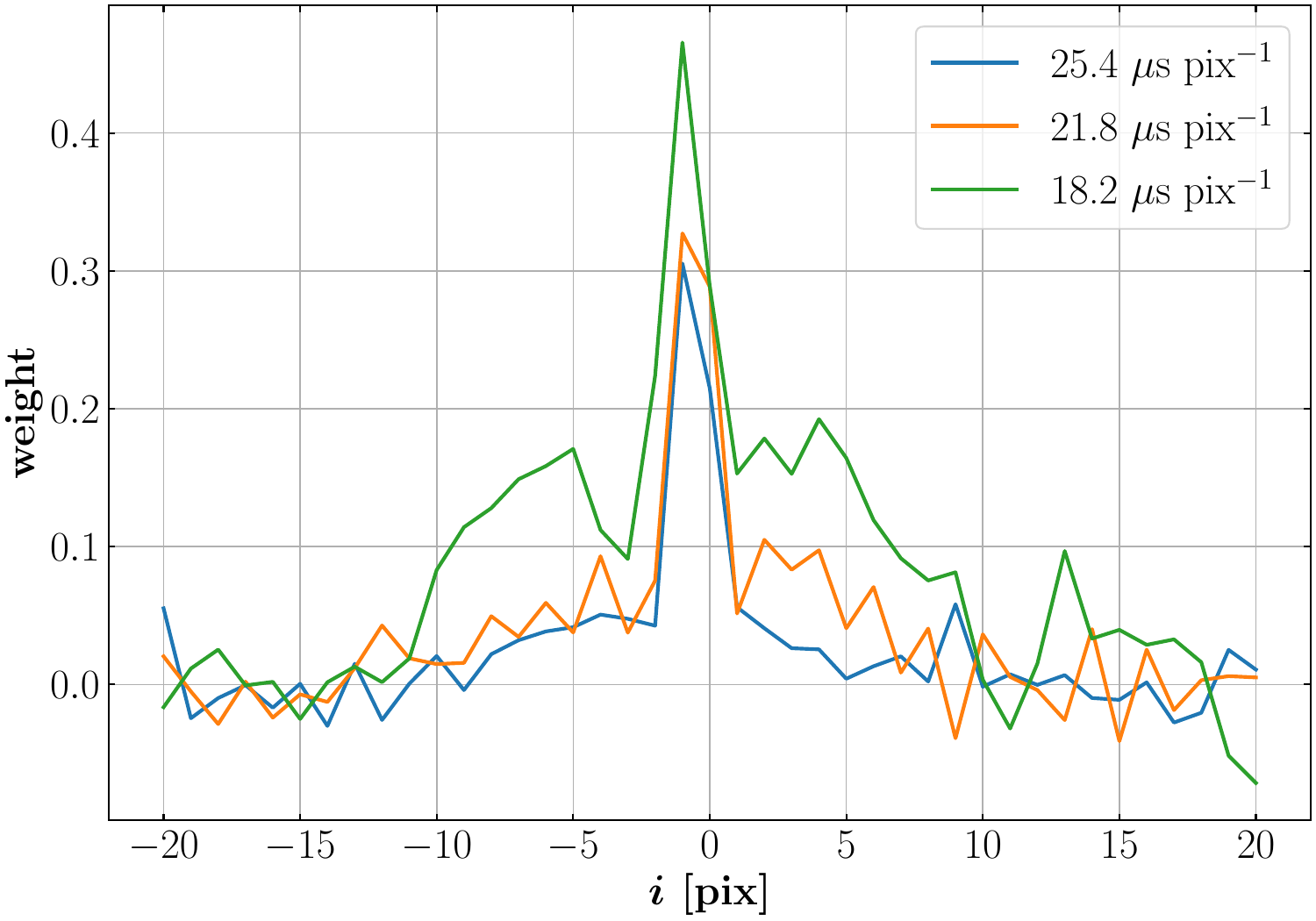}
    \caption{Weights calculated for $W=41$ for three exposures with different read times of 18.2, 21.8, and 25.4 $\mu$s pix$^{-1}$. There is still a peak at $i=-1$ and some oscillatory behavior in the tails, however the shape of the filter is now different for each exposure. The amplitude of the peak and where the oscillations occur are no longer consistent when comparing the exposures.}
    \label{fig:filter-shape}
\end{figure}

We show in Figure \ref{fig:weights} that when the read times are equivalent the shape of the filter is the same across multiple exposures. The amplitude of the peaks and the oscillations are consistent across the four exposures. When the read times are not equivalent, we find that the shape of the filter changes, though the weights always peak at $i~=~\{-1,0\}$. We show the shape of the filter for three different exposures with different integration times corresponding to total read times of 18.2, 21.8, and 25.4 $\mu$s pix$^{-1}$ in Figure \ref{fig:filter-shape}. We still observe oscillatory behavior in the tails, however, there are differences both in the amplitude of the peak and in the oscillations themselves, with the peaks and troughs of the oscillations occurring at different pixel shifts $i$ for the different read times. 

\begin{figure}
    \centering
    \includegraphics[width=\linewidth]{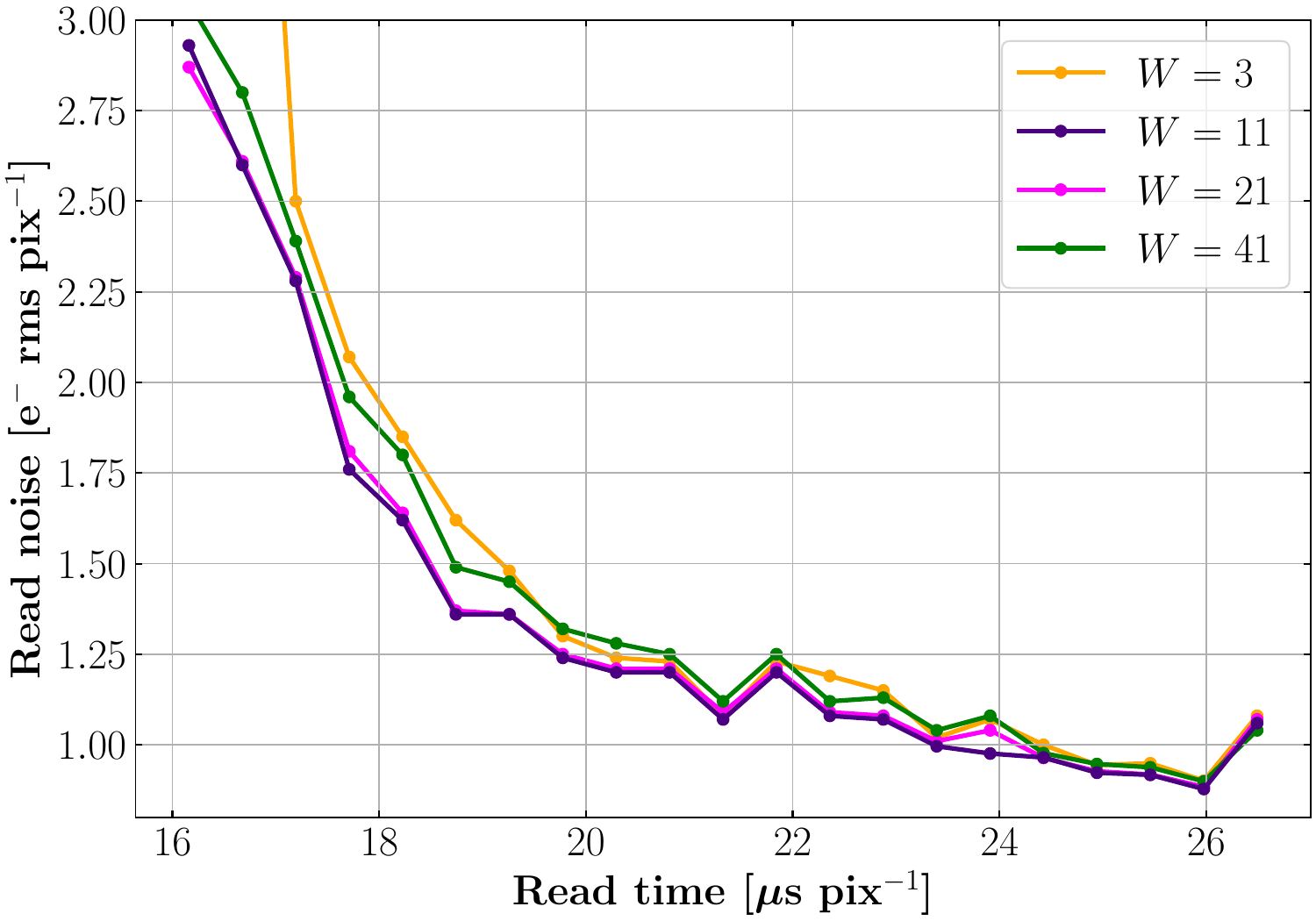}
    \caption{Effect of filter width $W$ on the measured read noise for different read times when the signal and reference level integration time are equal (discussed in the next section). While the four filter widths here produce similar results, we find empirically that the optimal value is $W=11$ (indigo line) and we adopt this value for all subsequent analysis in this paper.}
    \label{fig:rn-vs-rt}
\end{figure}

We show in Figure \ref{fig:rn-vs-rt} the read noise as a function of read time for four different filter widths, $W=3, 11, 21, 41$. The best read noise is obtained with filter widths of 11 and 21. We will use the smaller filter of $W=11$ for noise measurements in the remainder of this paper. In the next sections, we discuss additional strategies to further reduce the total read time while minimizing the noise.

\section{Optimizing Read Noise and Time}
\label{sec:readtime}
For astronomical imaging and spectroscopy where the signal is faint, the detector noise and the read time must be co-optimized for the target of interest \citep{2020SPIE11454E..1AD}. We investigate the pixel integration time, which dominates the total read time because the time set for each pixel to be sampled one at a time is the bottleneck that drives up the read time. Since both the read time and noise levels are affected by the integration time set for the signal and reference voltages, this parameter must be tuned carefully to reduce the readout time while minimizing the noise trade-off. Changing the number of signal and reset level samples that are read out is a key driver in the resulting total read time. 

\begin{figure}
   \centering
   \includegraphics[width=\linewidth]{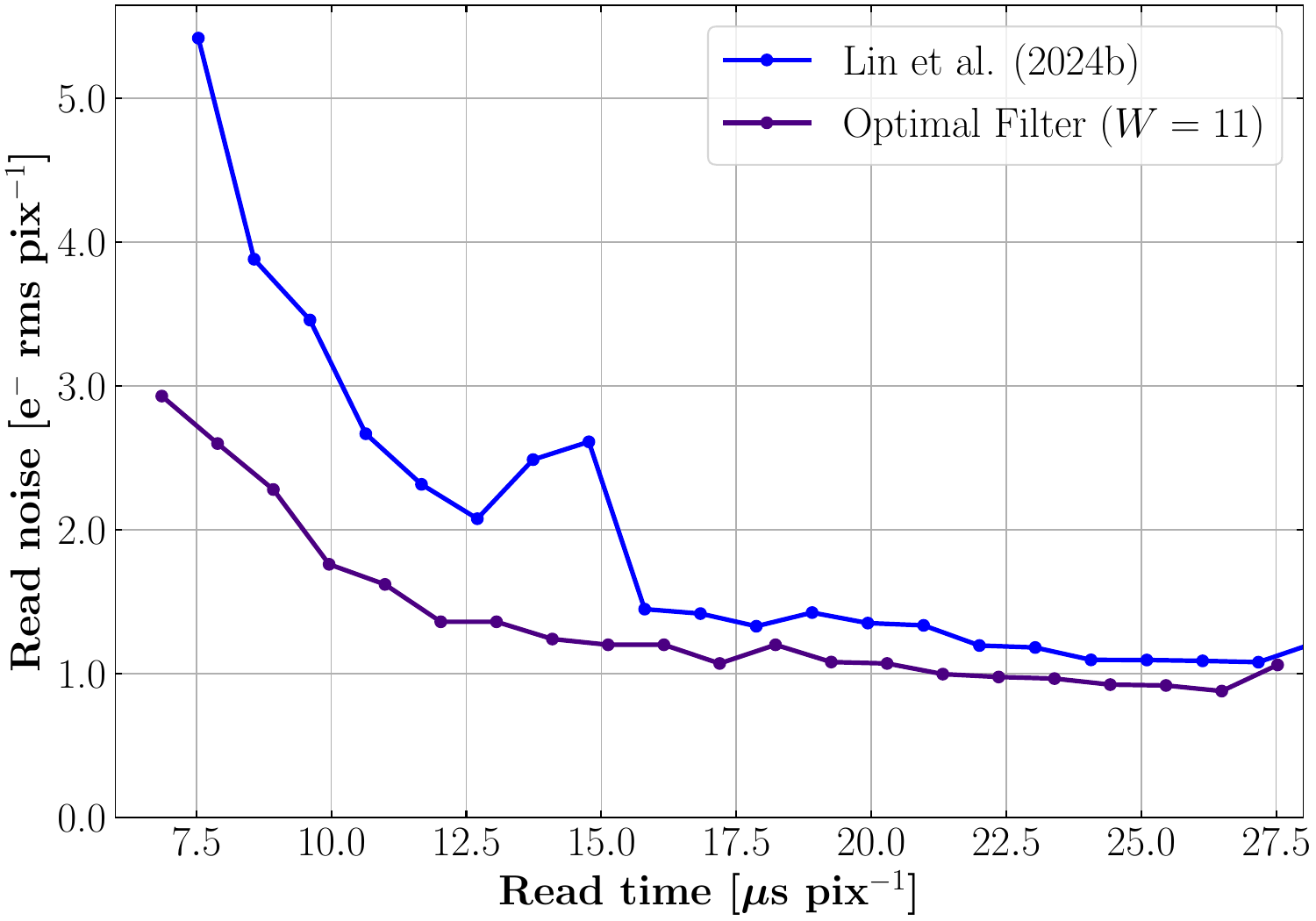}
   \caption{Read noise using the optimal filter with $W=11$ (indigo line) compared to the result from \citet{2024PASP..136i5002L} (blue line). Both results use the conventional CDS approach with equal integration times for the signal and reference level samples. By implementing only the optimal filter we are able to achieve significant reductions in read noise, especially at shorter read times.}
   \label{fig:optimal_vs_lin}
\end{figure}

We approach this first by starting with the conventional CDS mode where the number of signal samples is equal to the number of reset level samples. We start with short integration times for both the signal and reset levels and slowly ramp up each simultaneously to characterize any changes. The image pre-processing techniques described in Section \ref{sec:opportunities} are applied to each exposure to determine how fast we can read out while staying in the sub-electron noise regime. As shown in the indigo line in Figure \ref{fig:optimal_vs_lin} for a filter width of $W=11$, implementing an optimal filter alone allows us to achieve a lower read noise at all read times when compared to \cite{2024PASP..136i5002L} (blue line) operating under the same conditions. While the improvement at larger read times is minimal, the improvement at shorter read times is nearly double. Compared to DESI, which has a read time of 15 $\mu$s pix$^{-1}$ \citep{guy2022}, the read noise is improved by a factor of three to four and we are able to maintain sub-electron read noise for read times greater than 21.3 $\mu$s pix$^{-1}$. 

Because we can now obtain a reduced noise for the reference level, it is advantageous to spend more time measuring the signal than the reference level. In consequence, we can optimize jointly the number of samples for the signal and reference level in order to minimize the read noise for each read time value. In practice we performed several runs varying the integration time for the reference level from 1 $\mu$s to 10 $\mu$s and in the signal from 6 $\mu$s to 10 $\mu$s. The integration time of the signal was held constant during each run.

\begin{figure}
    \centering
    \includegraphics[width=\linewidth]{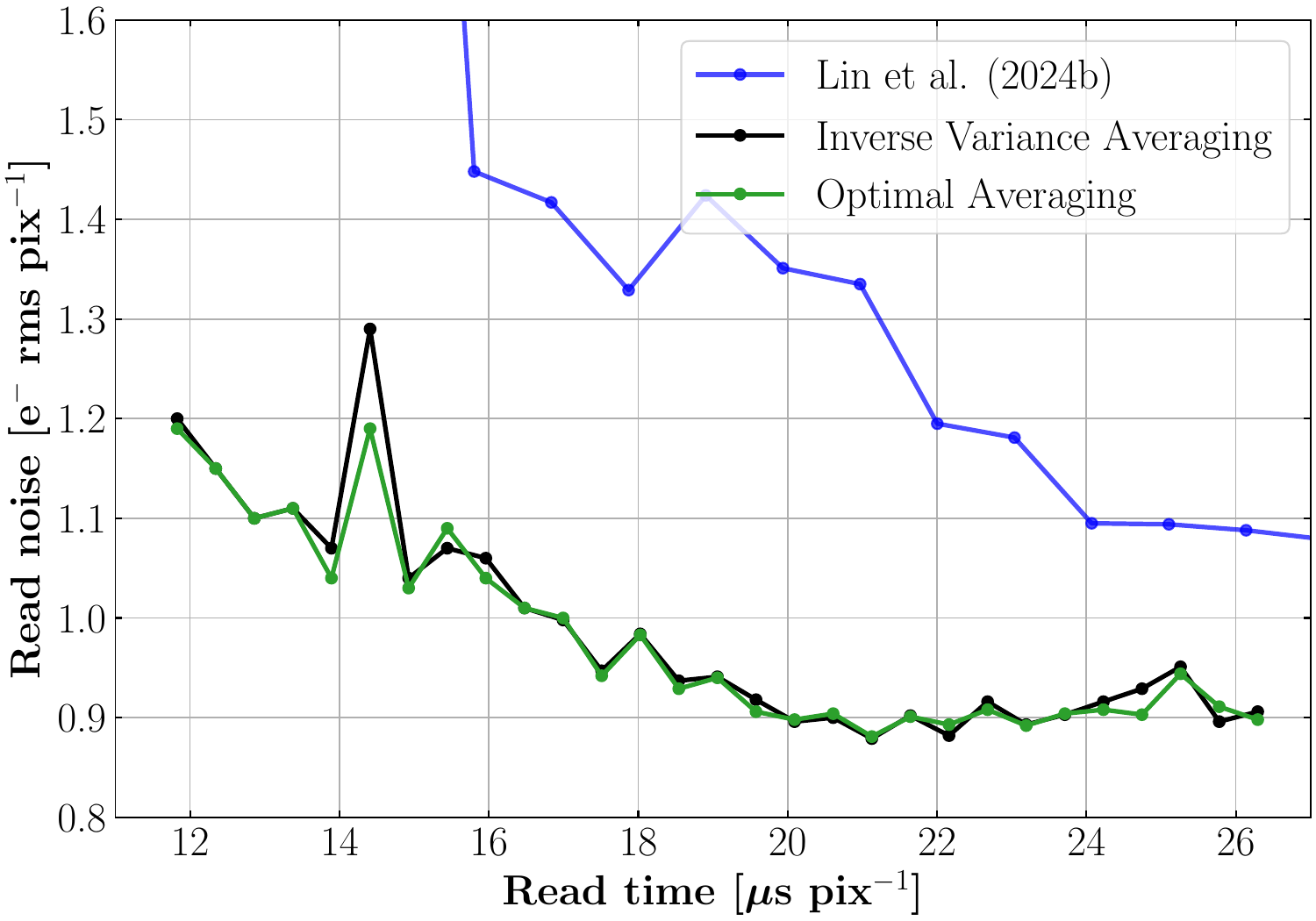}
    \caption{Read noise results from the runs where the reference level and signal integration times varied from 1 $\mu$s to 10 $\mu$s and 6 $\mu$s to 10 $\mu$s, respectively. For each run the signal integration time was held constant. We show the overall best read noise for the different integration times across all of the runs. The black line shows the results using inverse variance averaging and the green line shows the results using a cross-amplifier optimal weighting approach. The blue line shows the results from \cite{2024PASP..136i5002L} in Figure \ref{fig:optimal_vs_lin} for comparison. The optimal averaging minimizes the peak at 14.4 $\mu$s pix$^{-1}$ by 0.1 e$^-$.}
    \label{fig:averaging}
\end{figure}

We show in Figure \ref{fig:averaging} that by tuning the integration time for the signal and reference level we can achieve a better read noise with similar read time compared to the earlier case when using the conventional CDS approach. Though the conventional CDS approach can achieve shorter read times, the measurements have higher read noise because the number of signal (and reference level) samples decreases. In this approach, our best read noise of 0.88 e$^-$ rms pix$^{-1}$ is obtained for a read time of 21.1~$\mu$s pix$^{-1}$ (using a filter width $W=11$). At this read time, the signal and reference level integration times are 10 $\mu$s and 6 $\mu$s, respectively. We see that the shorter read times have higher read noise (because the number of reference level samples is small and therefore noisier), however we are able to obtain sub-electron read noise for read times greater than 17.5 $\mu$s pix$^{-1}$. There is a sharp peak at 14.4 $\mu$s pix$^{-1}$ associated with a combination of pattern and correlated noise. We show in Section \ref{optimalaveraging} that the correlated noise component can be reduced by optimally averaging the amplifiers.

\section{Optimal Amplifier Averaging} \label{optimalaveraging}
One potential optimization parameter space explored was the weights with which the image output from each amplifier is combined. The most straightforward approach for image stacking is simply taking the average of each pixel over the 16 amplifier images. However, this weights each amplifier equally, which may be sub-optimal if certain nodes contribute higher noise levels than others. If the combined image can be averaged using different weights, amplifiers that produce a larger pixel variance could be optimized to contribute less weight to the resulting averaged image. Assuming the amplifiers are completely uncorrelated, a simple scheme is to weight by the inverse variance. However, because correlated noise may be generated between amplifiers originating from either on the chip itself or in the readout electronics where each set of four amplifiers are read out using a discrete four-channel video processing board, we investigate optimally minimizing the noise contributions from each amplifier.

We apply a similar prescription outlined in \citet{10.1117/1.JATIS.11.1.011203} for calculating these optimal weights and the approach to the one considered for the optimal filter. We estimate the ideal weights by a least squares minimization where a constraint is now imposed such that the sum of the weights is equal to one, which was not necessary for the filter normalization because it was set by the comparison with the signal level. We adopt an approach using Lagrange multipliers to optimize these weights on each amplifier to minimize the noise of the combined image. When combining the images normally, each amplifier is given an equal weight of one and averaged over the number of amplifiers. To tune these weights $w_a$ and recover the optimally weighted average image, we apply the $\sum_a^{N_a} w_a = 1$ constraint. As a result, we have the Lagrangian function:
\begin{equation} \label{lagrangian}
    \mathcal{L} = \sum_p \left(\sum_{a,b,i} w_a w_b P_{a,i} P_{b,i} \right) + \lambda \cdot \left(\sum_a^{N_a} w_a - 1\right),
\end{equation}
where $w_a$ are the weights assigned for each amplifier $a$ and $P_i$ is a pixel value $i$ read at either amplifier $a$ or $b$. The expectation value of $\mathcal{L}$ is
\begin{equation} \label{expectationlagrangian}
    \langle \mathcal{L} \rangle = \sum_{a,b,i} w_a w_b \langle P_{a,i} P_{b.i} \rangle + \lambda \cdot \left(\sum_a^{N_a} w_a - 1\right)
\end{equation}
and is the function that we wish to minimize. In the trivial case where the pixel values in the same position read from two different amplifiers are uncorrelated $\langle P_{a,i} P_{b,i} \rangle = \delta_{ab} \text{Var}(P_a)$ where $\delta_{ab}$ is the Kronecker delta. In the case where amplifier correlation exists, then 
\begin{equation}
    \langle P_{a,i} P_{b,i} \rangle = \frac{1}{N_{\rm pix}} \sum_i^{N_{\rm pix}} P_{a,i} P_{b,i} 
\end{equation}
where $N_{pix}$ is the total number of pixels evaluated in the region of interest (e.g., overscan). Minimizing Eq. \ref{expectationlagrangian} then yields the optimal weights $w_a$ that gives the lowest combined noise.

Using this approach, we compute custom weights for a sample of exposures at varying integration time settings. This operation is carried out after images have been constructed from the optimal filtering of the reset level, subtraction, alignment, and bias (overscan) subtraction. The linear combination of each image from each amplifier with its corresponding weight is the optimally averaged image. We find that for integration times where the noise is at single-electron or under, the optimally weighted averaging gives at best a 0.03 e$^-$ rms pix$^{-1}$ reduction in the noise floor compared to weighting by the inverse variance, which assumes that each amplifier is independent and uncorrelated. The improvement is greater at peaks driven by pattern noise such as at 14.4 $\mu$s pix$^{-1}$, where a 0.1 e$^-$ drop is measured. This demonstrates that there is a detectable level of correlated noise between amplifiers that is partially removable with optimized averaging, and that the level of this correlation varies based on the integration time. The result of applying optimal averaging is shown as the green curve in Figure \ref{fig:averaging}. While the curve largely tracks that of the assumption of uncorrelated amplifiers, there are specific cases where high levels of correlation may be reduced by optimal averaging.

\begin{figure}
   \centering
   \includegraphics[width=\linewidth]{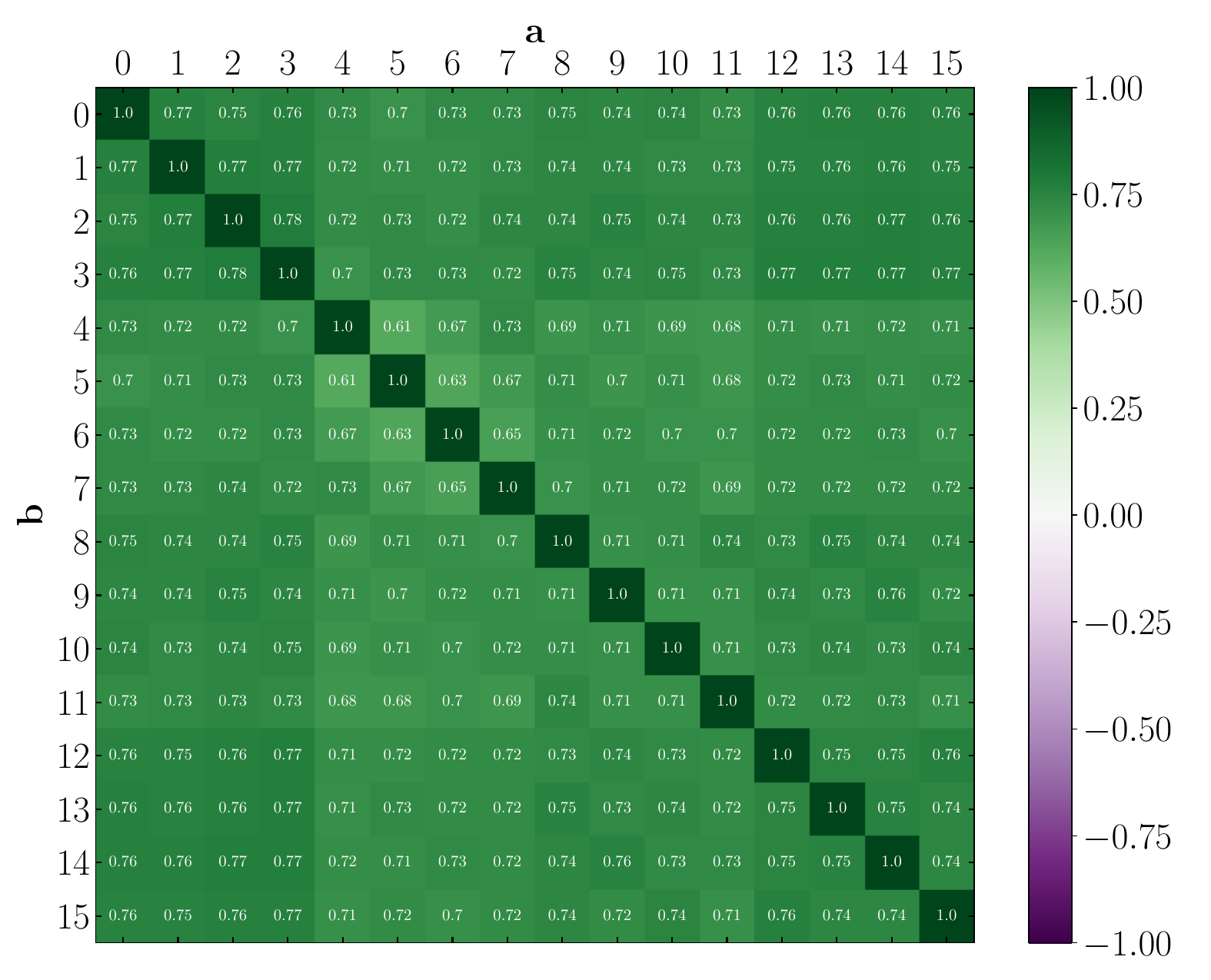}
   \caption{The measured correlation matrix between each amplifier pair $a$ and $b$ from 0 to 15 with the correlation indices corresponding to the colors in the color bar. The correlation coefficients in this matrix were measured at the peak in Figure \ref{fig:averaging} where the readout time was 14.4 $\mu$s pix$^{-1}$. Significant pickup noise is measured at shorter integration times and is reflected in strong correlated signal between amplifiers. In imaging, this effect manifests as electronic pattern noise.}
   \label{fig:corrmatrixpickupnoise}
\end{figure}

To understand and validate these results, we measured how correlated each amplifier was relative to each other to investigate common mode noise and verify whether noise decorrelation would be expected to yield a significant improvement, following the measurement presented in \citet{10.1117/1.JATIS.11.1.011203} which used a synchronized set of four Low-Threshold Acquisition (LTA) readout electronics \citep{2021JATIS...7a5001C}. For integration times shorter than 15 $\mu$s pix$^{-1}$, we see that a maxima in the green curve of Figure \ref{fig:averaging} is reduced but remains present. We compute the correlation coefficients at this integration time and represent this in the correlation matrix in Figure \ref{fig:corrmatrixpickupnoise}, where the colors correspond to how correlated each amplifier is to each other. We find that at this integration time with the noise maxima, there is significant correlated noise particularly between sets of neighboring amplifiers. Each of these groups share a video processing board, and suggests that the correlated component may be mitigated through improved readout electronics designed to read $\geq16$ channels that minimize pickup noise between signal lines. 

However, while the noise peak is flattened with optimal averaging, it is still at an elevated noise level because of the dominance of the fixed pattern noise on the image around certain integration times. This was also observed when reading out with conventional CDS where sample and reset integration times were equal. The read noise we measure includes both components of correlated and pattern noise, and in particular is dominated by the latter around integration times corresponding to the 14.4 $\mu$s pix$^{-1}$. This contrasts with most of the longer integration times in Figure \ref{fig:averaging} where the pattern noise component was non-detectable and a marginal noise improvement was realized from the removal of any trace correlated noise.

\begin{figure}
   \centering
   \includegraphics[width=\linewidth]{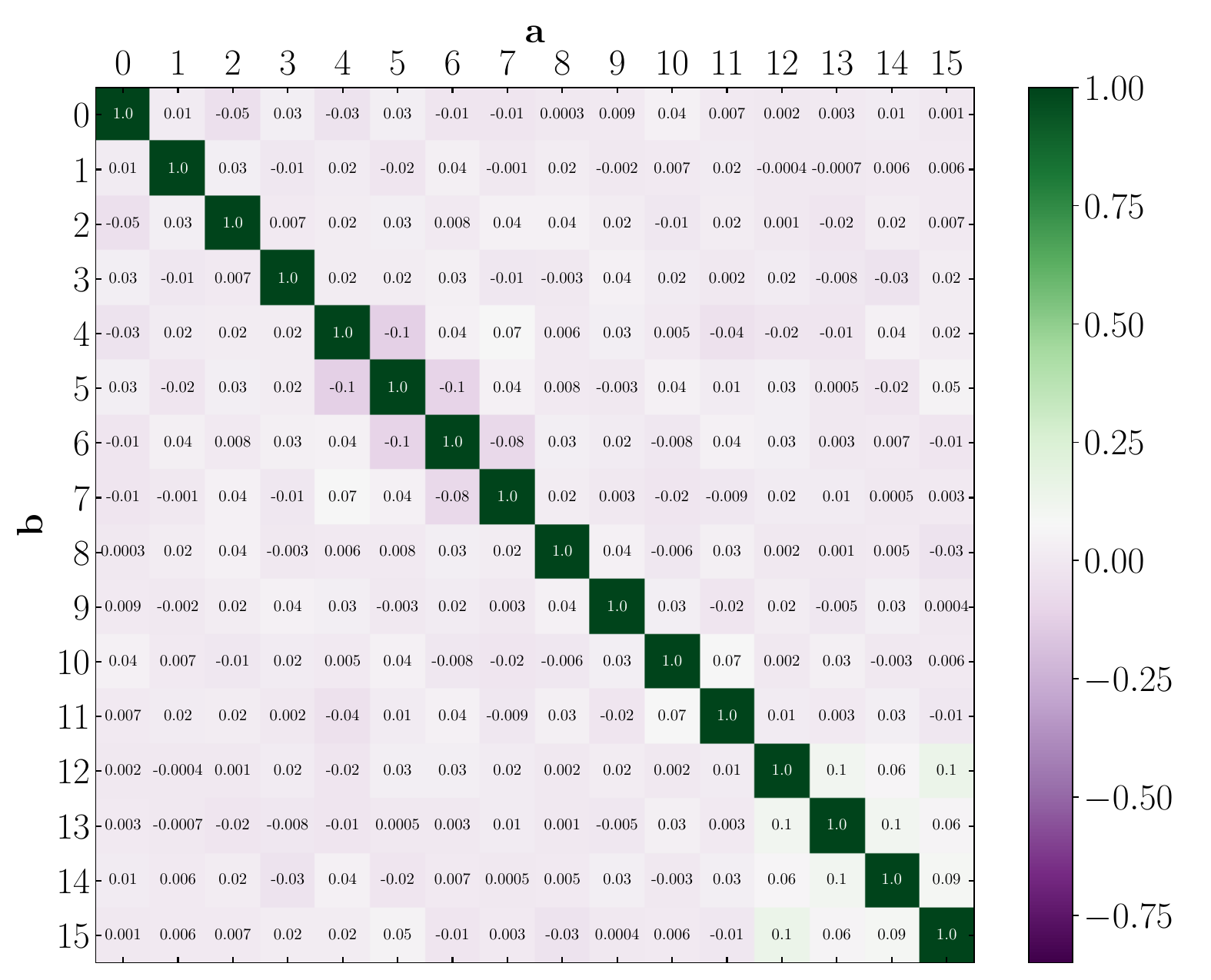}
   \caption{The correlation matrix measured between each amplifier pair $a$ and $b$ with the correlation indices illustrated by the colors, as in Figure \ref{fig:corrmatrixpickupnoise}, now for a different choice of integration time corresponding to 18.7 $\mu$s pix$^{-1}$. The level of correlation is weak between any given amplifier pair at this integration time, never exceeding an index of $|0.1|$ in magnitude. No pattern noise pickup was detected in this integration time regime.}
   \label{fig:correlationmatrix}
\end{figure}

In the regime where pattern noise is not observed ($\gtrsim 15$ $\mu$s pix$^{-1}$), we find the correlation to be weak between any pair of amplifiers, with all coefficients under $|0.1|$ as shown in Figure \ref{fig:correlationmatrix}. The highest correlation indices are associated with the same cluster of four amplifiers between 12 and 15 that had the highest coefficients in Figure \ref{fig:corrmatrixpickupnoise}. However, this correlation originating from a single video board is comparatively weak and demonstrates why the optimal weights for averaging the amplifiers are nearly equivalent to that of using simple inverse variance weights where no correlation is assumed and where no pattern noise is evident. We note that the magnitude of correlated noise is highly dependent on the readout electronics, and we have shown that outside integration times affected by pattern noise, the Hydra electronics have minimal impact on contributing unwanted correlated noise.

\section{Summary} \label{sec:summary}
The floating-gate amplifier coupled with the serial amplifier chain structure of the MAS CCD has enabled new readout and processing techniques to be fully realized, yielding sub-electron noise performance at 17.5~$\mu$s~pix$^{-1}$. Using readout electronics based on the DESI CCD controller, we have demonstrated the ability to exploit the non-destructive floating-gate amplifier to reduce the number of resets to once per line, which in turn provides the opportunity to apply filtering techniques to the reset level, thereby reducing the read noise by removing the noise components present in the reset level in each line. Manual processing of the CDS operation has also enabled the ability to tune the integration times of the reset and signal levels, which has driven an increase in the readout speed. We note that all of the approaches presented here can be generalized for any non-destructive repetitively sampling detector, such as the original Skipper CCD where charge is measured at a single floating-gate sense node.

At a read time of 21.1 $\mu$s pix$^{-1}$, our lowest demonstrated noise using these techniques was 0.88~e$^-$~rms~pix$^{-1}$. When determining the optimal filter we found that a filter width of $W=11$ produced the lowest read noise. Although the optimal filter width is independent of read time, we found that the shape of the filter does change depending on the read time. The location of the peak was consistent, however the amplitude of the peak and the oscillations themselves were different for different read times.

We explored different integration time schemes by changing the number of reference level samples measured during read out while the number of signal samples remained constant. While the read time of 21.1 $\mu$s pix$^{-1}$ produced the lowest read noise we have experimentally achieved, we consistently reach a sub-electron read noise for read times greater than 17.5 $\mu$s pix$^{-1}$. This mode also produced lower read noise at all read times compared to integration times where the number of signal and reference level samples were equal. 

We also found that for most integration times, there was minimal cross-amplifier noise. By averaging the 16 amplifier images using optimal weights and comparing with averaging assuming uncorrelated amplifiers, we found that the noise measurements are mostly consistent. In these cases, the correlations between amplifiers are weak, with all coefficients less than $|0.1|$. With these weak correlations, the optimal averaging, compared to the inverse variance averaging, gives at most a 0.03 e$^-$ rms pix$^{-1}$ reduction in the noise. However, we also find that at certain specific integration times where there is correlated noise, a larger improvement of nearly 0.1 e$^-$ may result from optimally averaging the information from the 16 amplifiers. While we expect that the particular features we observed will be specific to the readout electronics used, we have demonstrated that the optimal averaging of multiple measurements can provide a more precise measurement of the pixel charge.

The readout and optimization techniques we exhibited in this work lays the foundation for a more realistic large-format science-grade MAS CCDs to be readied for on sky deployment. This next iteration of MAS CCDs under fabrication would feature not only a larger active area (e.g., 4k $\times$ 4k), but also an increased number of output amplifiers per quadrant. They would be designed to allow native operation in a differential readout mode where each channel provides a signal or reference level output, which exploits the approach that we have experimentally demonstrated with our prototype. We expect to test and characterize these devices using a mature version of the Hydra readout electronics that we are designing specifically for the MAS CCD and CCDs that feature a large number of outputs.

As future spectroscopic surveys probe fainter, more distant objects, or require higher spectral-resolution, the need for an extremely low-noise detector becomes increasingly important. Potential ground-based high-redshift spectroscopic surveys for precision cosmology such as Spec-S5 and planned targeted spectroscopy of faint Earth-like exoplanets by HWO depend on sub-electron and photon-counting detector capabilities with high quantum efficiency from the optical to near-IR. Harnessing the readout strategies enabled by floating-gate amplifiers that give a 30\% advantage in readout time with simultaneous noise reduction, the prototype MAS CCD already represents a solution for next generation spectroscopic surveys from the ground. As a potentially competitive and attractive solution for the fast ($<14$ $\mu$s pix$^{-1}$), photon-counting demands of HWO, our demonstration motivate advanced design efforts aimed at optimizing single-sample noise at the device level for future studies.

\begin{acknowledgments}
Acknowledgments: This research was supported by the Laboratory Director's Research and Development (LDRD) Program at Lawrence Berkeley National Laboratory under Contract DE-AC02-05CH11231. The multi-amplifier sensing (MAS) CCD was developed as a collaborative endeavor between the Lawrence Berkeley National Laboratory and the Fermi National Accelerator Laboratory. Funding for the design and fabrication of the MAS device used in this work came from a combination of sources, including the DOE Quantum Information Science (QIS) initiative, the DOE Early Career Research Program, and the Laboratory Directed Research and Development Program at Fermi National Accelerator Laboratory under Contract No. DE-AC02-07CH11359. We thank the CCD group at Fermilab for providing the device that was used in this study.
\end{acknowledgments}

\vspace{5mm}

\software{astropy \citep{2013A&A...558A..33A,2018AJ....156..123A},
          scipy \citep{2020SciPy-NMeth}, 
          Source Extractor \citep{1996A&AS..117..393B, Barbary2016}
          }

\bibliography{final}{}
\bibliographystyle{aasjournal}
\end{document}